# Quadrupole collectivity in Si isotopes around $N = 20$. *

R.R. Rodríguez-Guzmán, J.L. Egido and L.M. Robledo

Departamento de Física Teórica C-XI, Universidad Autónoma de Madrid,
28049-Madrid, Spain.

The angular momentum projected Generator Coordinate Method using the quadrupole moment as collective coordinate and the Gogny force as the effective interaction is used to describe the properties of the ground state and low-lying excited states of the neutron rich light nuclei $^{32,34,36}Si$. It is found that the ground state of the nucleus $^{34}Si$ is spherical. However, this is not only due to the $N = 20$ shell closure as the ground state of $^{34}Si$ contains a significant amount of the intruder $f_{7/2}$ neutron orbital. On the other hand, rather good agreement with experimental data for many observables is obtained.

PACS numbers: 21.60.Jz, 21.60.-n, 21.10.Re, 21.10.Ky, 21.10.Dr, 27.30.+t

## 1. Introduction

In nuclear physics the mean field approximation is always the first step to understand the properties of the ground and lowest-lying excited states. The mean field approximation provides the concept of magic numbers as well as the concept of spontaneous symmetry breaking. For nuclei with proton and/or neutron numbers close to the magic ones one expect symmetry conserving (i.e. non superconducting and spherical) ground states. On the other hand, for nuclei away from the magic configurations one expect strong symmetry breaking and the appearance of deformed ground states that generate bands (as the rotational bands).

The experimental studies of light nuclei away from the stability line $N = Z$ seem to imply that for those (usually neutron rich) nuclei some of the properties associated to magic numbers are not preserved. The most striking example is the experimental evidence towards the existence of quadrupole deformed ground states in the neutron-rich nuclei around the magic number $N = 20$. In addition, the extra binding energy coming from deformation

---

can help to extend thereby the neutron drip line in this region far beyond what could be expected from spherical ground states. Among the variety of available experimental data, the most convincing evidence for a deformed ground state in the region around $N = 20$ is found in the $^{32}Mg$ nucleus where both the excitation energy of the lowest lying $2^+$ state [1] and the $B(E2, 0^+ \to 2^+)$ transition probability [2] have been measured. Both quantities are fairly compatible with the expectations for a rotational state. At the mean field level, the ground state of $^{32}Mg$ is spherical. However, when the zero point rotational energy correction (ZPRE) is considered, the energy landscape as a function of the quadrupole moment changes dramatically and $^{32}Mg$ becomes deformed [3, 4, 5, 6, 7, 8, 9].

A more careful analysis of the energy landscape including the ZPRE correction reveals that, in fact, there are two coexistent configurations (prolate and oblate) with comparable energy indicating thereby that configuration mixing of states with different quadrupole intrinsic deformation has to be considered. Therefore, an Angular Momentum Projected Generator Coordinate Method (AMPGCM) calculation with the quadrupole moment as collective coordinate is called for. We have applied this method in Ref. [10] to the study of the nuclei $^{30-34}Mg$. The Gogny force [11] (with the D1S parameterization [12]) has been used in the calculations. We have obtained prolate ground states for $^{32-34}Mg$ indicating that the $N = 20$ shell closure is not preserved for the $Mg$ isotopes. Moreover, a good agreement with the experimental data for the $2^+$ excitation energies and $B(E2)$ transition probabilities was obtained. The purpose of this paper is to extend those calculations to the study of the nuclei $^{32-36}Si$ which are obtained by adding two protons to the $^{30-34}Mg$ nuclei. These nuclei have recently being studied experimentally [13] and their ground states seem to be spherical and therefore preserve the $N = 20$ shell closure. Therefore, our intention is to check whether our method and interaction is not only able to predict a deformed ground state in $^{32}Mg$ (Z=12) but also to predict an spherical one in $^{34}Si$ (Z=14).

## 2. Theoretical framework

The angular momentum projected Generator Coordinate Method (AMP-GCM) with the mass quadrupole moment as generating coordinate is used as the theoretical framework. As we restrict ourselves to axially symmetric configurations, we use the following ansatz for the $K = 0$ wave functions of the system

$$\left|\Phi_\sigma^I\right\rangle = \int dq_{20} f_\sigma^I(q_{20}) \hat{P}_{00}^I \left|\varphi(q_{20})\right\rangle \qquad (1)$$

In this expression $|\varphi(q_{20})\rangle$ is the set of axially symmetric (i.e. $K = 0$) Hartree-Fock-Bogoliubov (HFB) wave functions generated by constraining the mass quadrupole moment to the desired values $q_{20} = \langle\varphi(q_{20})| z^2 - 1/2(x^2+y^2) |\varphi(q_{20})\rangle$. The quasiparticle operators associated to the intrinsic wave functions $|\varphi(q_{20})\rangle$ have been expanded in a Harmonic Oscillator (HO) basis containing 10 major shells and with equal oscillator lengths to make the basis closed under rotations [14]. The operator

$$\hat{P}^I_{00} = \frac{(2I+1)}{8\pi^2} \int d\Omega d^I_{00}(\beta) e^{-i\alpha \hat{J}_z} e^{-i\beta \hat{J}_y} e^{-i\gamma \hat{J}_z} \qquad (2)$$

is the usual angular momentum projector with the $K = 0$ restriction [15] and $f^I_\sigma(q_{20})$ are the "collective wave functions" solution of the Hill-Wheeler (HW) equation

$$\int dq'_{20} \mathcal{H}^I(q_{20}, q'_{20}) f^I_\sigma(q'_{20}) = E^I_\sigma \int dq'_{20} \mathcal{N}^I(q_{20}, q'_{20}) f^I_\sigma(q'_{20}). \qquad (3)$$

In the equation above we have introduced the projected norm

$$\mathcal{N}^I(q_{20}, q'_{20}) = \langle\varphi(q_{20})| \hat{P}^I_{00} |\varphi(q'_{20})\rangle \qquad (4)$$

and the projected hamiltonian kernel

$$\mathcal{H}^I(q_{20}, q'_{20}) = \langle\varphi(q_{20})| \hat{H} \hat{P}^I_{00} |\varphi(q'_{20})\rangle. \qquad (5)$$

As the generating states $\hat{P}^I_{00} |\varphi(q_{20})\rangle$ are not orthogonal, the "collective amplitudes" $f^I_\sigma(q_{20})$ cannot be easily interpreted. Instead, one usually introduce [16] the "collective" amplitudes

$$g^I_\sigma(q_{20}) = \int dq'_{20} f^I_\sigma(q'_{20}) \mathcal{N}^I(q_{20}, q'_{20})^{1/2*} \qquad (6)$$

which are orthonormal $\int dq_{20} (g^I_\sigma)^*(q_{20}) g^{I'}_{\sigma'}(q_{20}) = \delta_{I,I'}\delta_{\sigma,\sigma'}$ and therefore their module squared has the meaning of a probability.

The $B(E2)$ transition probabilities are computed using the AMPGCM wave functions as

$$B(E2, I_i \to I_f) = \frac{e^2}{2I_i+1} |\int dq^i_{20} dq^f_{20} f^{I_f *}_{\sigma_f}(q^f_{20}) \langle I_f q^f_{20} || \hat{Q}_2 || I_i q^i_{20}\rangle f^{I_i}_{\sigma_i}(q^i_{20}) |^2 \qquad (7)$$

As we are using the full configuration space no effective charges are needed. Further details on the computational procedure can be found in Ref [10].

### 3. Discussion of the results

In Figure 1 we show the HFB potential energy surfaces (PES) for the three nuclei considered (dashed line) along with the projected energies $E^I = \mathcal{H}^I(q_2,q_2)/\mathcal{N}^I(q_2,q_2)$ for $I = 0, 2, 4, 6$ and 8 (full lines). The HFB PES show a very pronounce spherical minimum in $^{34}Si$, a shallow spherical minimum in $^{36}Si$ and, finally, a very swallow oblate minimum in $^{32}Si$. These results are rather different from the ones obtained for the corresponding $Mg$ isotopes with the same neutron number [10]. The reason for such differences is that the $Si$ isotopes have the proton $d_{5/2}$ shell completely occupied and therefore, the up-slopping $K = 5/2$ level of the proton $d_{5/2}$ orbital prevents the appearance of prolate deformation. At the mean field level we can say that the $Si$ isotopes around $N = 20$ retain the closed shell properties associated to the $N = 20$ shell closure. However, as it has already been recognized, the effect of the restoration of the rotational symmetry in the HFB PES can be very important and it can substantially modify the conclusions extracted from the mean field results. The angular momentum projected (AMP) PES of Figure 1 show in the three nuclei considered two minima, one prolate and the other oblate, for all the angular momentum considered. However, the prolate and oblate minima lie rather close in energy in almost all the cases and therefore, we expect strong mixing of the two configurations. Taking into account this fact and also that the amount of mixing is not only determined by the AMP PES but also by the "collective inertia" the amount of mixing can only be disentangled when a configuration mixing calculation is carried out.

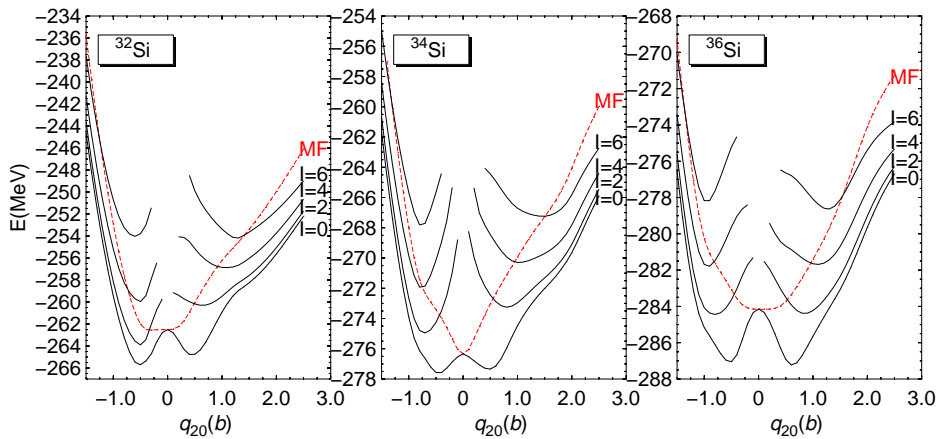

Fig. 1. The HFB potential energy surface (PES) for the three nuclei considered (dashed line) together with the projected energies for $I = 0, \ldots, 8$ (full lines).

We have also carried out configuration mixing (in the framework of the GCM) calculations taking into account angular momentum projected wave functions and the quadrupole constrained HFB wave functions as generating intrinsic states. The solution of the HW equation provides us with energies $E^I_\sigma$ and collective wave functions $f^I_\sigma(q_{20})$. As mentioned in section 2 the amplitudes $f^I_\sigma$ are not well suited for a physical interpretation and therefore one considers the collective amplitudes $g^I_\sigma(q_{20})$ instead. In terms of these amplitudes one can defined an "averaged" intrinsic quadrupole moment $(\overline{q}_{20})^I_\sigma = \int dq_{20}\, q_{20} \left|g^I_\sigma(q_{20})\right|^2$ for each of the AMPGCM solutions.

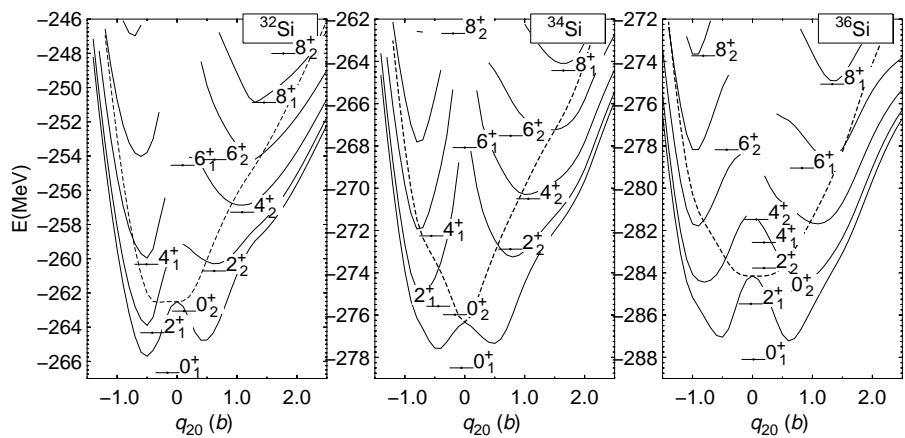

Fig. 2. The HFB potential energy surface (PES) for the three nuclei considered (dashed line) together with the projected energies for $I = 0, 2, 4, 6$ and $8$ (full lines).

In Figure 2 we show, along with the angular momentum projected PES (plotted again to guide the eye), the energies $E^I_\sigma$ solution of the HW equation for $\sigma = 1$ and 2 and angular momenta $I = 0, 2, 4, 6,$ and $8$. Each energy has been placed at a value of the quadrupole moment corresponding to the "averaged" intrinsic quadrupole moment $(\overline{q}_{20})^I_\sigma$. From figure 2 we see that in $^{32}Si$ the two $0^+$ states shown are spherical, the $2^+_1$ and $4^+_1$ are oblate whereas the $2^+_2$, $4^+_2$, $6^+_2$ and $8^+_1$ are prolate deformed states. The $6^+_2$ state is almost degenerate in energy with the nearly spherical $6^+_1$ state and this could explain the anomalous quadrupole moment of the former state. A nearly identical pattern is also seeing in the nucleus $^{34}Si$ except for the position of the $0^+_2$ state that lies below the $2^+_1$. In the nucleus $^{36}Si$ the ground state is spherical and a prolate deformed $0^+_2$ state is obtained. The $2^+$ and $4^+$ states are all of them almost spherical whereas the $6^+_1$ and $8^+_1$ are prolate and the $6^+_2$ and $8^+_2$ states are oblate. We observe that in the three nuclei considered the ground state is spherical due to the configuration mixing between the prolate and oblate minima. The rest of the spectra shows rather unclear patterns that can only be elucidated by looking at the "collective amplitudes" $g_\sigma^I(q_{20})$.

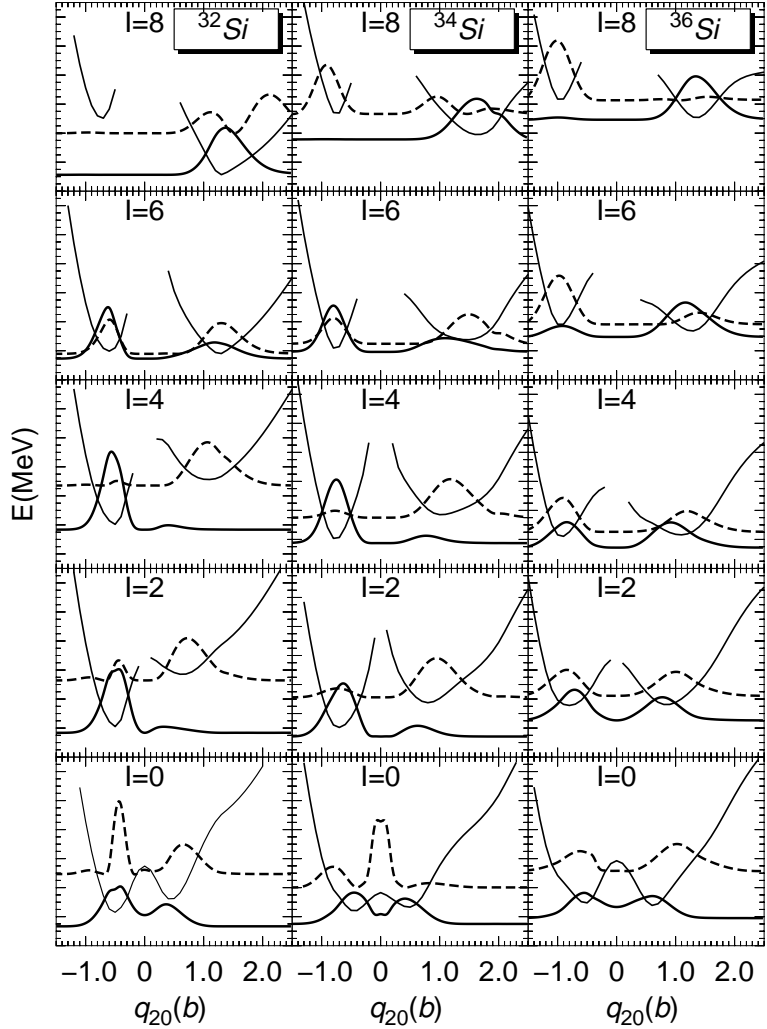

Fig. 3. The collective amplitudes $|g_\sigma^I(q_{20})|^2$ (thick lines) for $\sigma = 1$ (full) and 2 (dashed) and spin values of $I = 0\hbar, \ldots, 8\hbar$ for the nuclei $^{32}Si$, $^{34}Si$ and $^{36}Si$. The projected energy curve for each spin is also plotted (thin line). The y-axis scales are in energy units and span an energy interval of 13 MeV. The collective amplitudes have also been plotted against the energy scale after a proper scaling and shifting, that is, the quantity $E_\sigma^I + 15 \times |g_\sigma^I(q_{20})|^2$ is the one actually plotted.

In figure 3 the collective wave functions squared $\left|g_\sigma^I(q_{20})\right|^2$ for the two lowest solutions $\sigma = 1$ and 2 obtained in the AMP-GCM calculations are depicted. We also show in each panel the projected energy for the corresponding spin. We observe that the $0_1^+$ ground state wave functions of the three nuclei contain significant admixtures of both the prolate and oblate configurations. Looking into the single particle diagram of $^{34}Si$ (not shown here) it is observed that the neutron $f_{7/2}$ shell, that is empty at $q_{20} = 0$ and 5.78 MeV higher than the $d_{3/2}$, crosses the Fermi surface at the deformations $q_{20} = -0.5b$ in the oblate side (the $K = 7/2$ level) and $q_{20} = 0.5b$ in the prolate one (the $K = 1/2$ level). Therefore, the intrinsic states with $q_{20} > 0.5b$ and $q_{20} < -0.5b$ have (at least) two particles in the intruder $f_{7/2}$ orbital. As the ground state collective amplitude in $^{34}Si$ (and in the other two nuclei) is not negligible in the intervals $q_{20} > 0.5b$ and $q_{20} < -0.5b$ we conclude that the ground state of the three nuclei contain significant amounts of the intruder $f_{7/2}$ neutron configuration in spite of being spherical. To make the argument more quantitative we have computed the quantity $1 - \int_{-0.5}^{0.5} dq_{20} \left|g_\sigma^{I=0}(q_{20})\right|^2$ that gives us an idea of the percentage of the "collective" amplitude outside the $q_{20}$ interval between -0.5 and 0.5 $b$ (i.e. the interval where the intrinsic states contain the intruder $f_{7/2}$ configuration). This quantity is roughly 0.35 for $\sigma = 1$ in the three nuclei considered and it corresponds roughly to 0.7 particles in the neutron $f_{7/2}$ sub-shell. Therefore, we conclude that the ground state of $^{34}Si$ does not correspond to a pure configuration with $N = 20$ acting as a closed shell. Similar conclusions have been obtained in shell model calculations [17] and also in the Shell Model Monte Carlo calculations of [18].

The collective amplitude of the $0_2^+$ excited state of $^{34}Si$ is rather unusual, being concentrated around $q_{20} = 0$. This unusual behavior probably is the responsible for the low excitation energy of this state (see below). In the $^{32}Si$ and $^{34}Si$ nuclei and $I = 2$ and 4 the collective amplitudes are well located inside the corresponding prolate and oblate minima. For $I = 6$ we have accidentally coexisting prolate and oblate minima and for $I = 8$ the lowest minima becomes the prolate one. In the nucleus $^{36}Si$ and $I = 2$ and 4 we have shape coexistence between the prolate and oblate minima and therefore the collective amplitudes are equally distributed among them; the net result being that these states are, on the average, spherical. For $I = 6$ and 8 the prolate minimum becomes dominant and the collective amplitudes are localized around the minima.

Now, we want to compare our results with the available experimental data. The first piece of information we can compare to is the two neutron and two proton separation energies. The two neutron separation energies $(S(2n))$ for $^{34}Si$ and $^{36}Si$ are 11.83 MeV and 9.6 MeV respectively. These numbers have to be compared with the experimental data [19]: 12.018 MeV and 8.587 MeV, respectively. We observe a reasonable good agreement with experiment for both nuclei. Concerning the two proton separation energies ($S(2p)$) we have used the previous results [10] for the magnesium isotopes to obtain 28.70 MeV, 32.72 MeV and 36.26 MeV for $^{32-36}Si$, respectively. The corresponding experimental values [19] are 29.777 MeV, 33.737 MeV and 35.430 MeV. The agreement between theory and experiment is again rather reasonable.

|         | Energies (MeV) |      |      | Exp.  | $B(E2)e^2fm^4$ |       |       | Exp.     |
|---------|----------------|------|------|-------|----------------|-------|-------|----------|
|         | a              | b    | c    | a     | d              | e     | f     | d        |
| $^{32}Si$ | 2.34           | 3.60 | 5.95 | 1.941 | 82.18          | 0.01  | 28.00 | 113 ± 33 |
| $^{34}Si$ | 2.92           | 2.52 | 5.62 | 3.327 | 108.99         | 50.41 | 76.92 | 85±33    |
| $^{36}Si$ | 2.63           | 3.24 | 4.32 | 1.399 | 211.55         | 65.44 | 48.81 | 193 ± 59 |

Table 1. Calculated and experimental results for excitation energies (in MeV) and $B(E2, 0^+_{\sigma_1} \to 2^+_{\sigma_2})$ transition probabilities (in $e^2fm^4$). The columns marked a, b and c correspond to $0^+_1 - 2^+_1$, $0^+_1 - 0^+_2$ and $0^+_1 - 2^+_2$ respectively. The ones marked d,e and f correspond to $0^+_1 - 2^+_1$, $0^+_2 - 2^+_1$ and $0^+_1 - 2^+_2$ respectively.

In table 1 the energy splittings between different states and the $B(E2)$ transition probabilities among some of them are compared with the available experimental data. Concerning the $B(E2, 0^+_1 \to 2^+_1)$ transition probabilities we find a quite good agreement with the experiment. In fact, our calculation reproduces the increase of the $B(E2)$ values in going from $^{34}Si$ to $^{36}Si$. The other transition probabilities are much smaller than the $0^+_1 \to 2^+_1$ ones except in $^{34}Si$. On the other hand, the theoretical values for the $2^+_1$ excitation energies agree reasonably well with the experimental data for $^{32}Si$ and $^{34}Si$ but shows a discrepancy of 1.2 MeV in the case of $^{36}Si$. This discrepancy can be attributed to many sources, like not taking into account triaxial intrinsic configurations or not dealing with pairing correlations in a "beyond mean field" framework. Another source for the discrepancy could be related to the fact that ours is a calculation of the Projection After Variation (PAV) type instead of the more complete projection before variation (PBV). Usually, the PBV method yields to rotational bands with moments of inertia larger than the PAV ones [20, 21]. In Ref [10] we estimated the effect of considering PBV in calculations similar to the ones performed here but for some deformed magnesium isotopes. The conclusion was that the moment of inertia gets enhanced by a factor 1.4 and therefore the excitation energies have to be quenched by a factor 0.7. Unfortunately, the method used in [10] for the mentioned estimation can not be applied here as we are dealing with spherical ground states as well as with nearly spherical excited states.

On the other hand, shell model calculations [13, 17, 22] predict a second $0_2^+$ excited state in $^{34}Si$ with an excitation energy (2 MeV in [13], 3 MeV in [17] and 2.6 MeV in [22]) which is below the excitation energy of the $2_1^+$ state. In the experimental work of Ref [13] such excited state was not observed but it was argued that with the given experimental setup it was quite difficult to detect it. However, in the experiment reported in [22] the mentioned excited $0^+$ state was observed at a tentative excitation energy of 2.1 MeV. In our calculations we also observe a second $0^+$ state in $^{34}Si$ (spherical in character) which lies at an excitation energy of 2.52 MeV. Our predicted $B(E2)$ transition probability $0_2^+ \to 2_1^+$ is $50.41 e^2 fm^2$ which is a factor four smaller than the one obtained in the shell model calculations of [13, 17] and a factor of six smaller than the predicted value of [22]. In [22] an experimental indirect estimate of the $B(E2)$ transition probability of 444(210) $e^2 fm^4$ is given. At this point it has to be recalled that the $B(E2)$ transition probabilities are more sensitive to the collective wave functions than the excitation energies. Therefore, a relatively small change in the collective wave functions can alter drastically the transition probabilities. In our case the collective amplitude of the $0_2^+$ state (see Fig. 3) is concentrated around the spherical intrinsic configuration (result which is no consistent with the shell model expectations of [17, 22]). This spherical intrinsic configuration has zero pairing correlations in our HFB calculations and therefore it can be expected that the inclusion of pairing correlations in a "beyond mean field" framework can modify the properties of the quadrupole dynamics around this point. To elucidate this and other deficiencies a simultaneous projection on angular momentum and particle number (like to one performed in [8] with the Skyrme interaction) is called for.

## 4. Conclusions

In conclusion, we have performed angular momentum projected Generator Coordinate Method calculations with the Gogny interaction D1S and the mass quadrupole moment as generating coordinate in order to describe quadrupole collectivity in the nuclei $^{32}Si$, $^{34}Si$ and $^{36}Si$. We obtain a spherical ground state in $^{34}Si$ (in opposition to the nucleus $^{32}Mg$) but it contains significant admixtures of the intruder configuration with two neutrons in the $f_{7/2}$ orbital. We also find in $^{34}Si$ that the first excited state is a $0^+$ with an excitation energy of 2.5 MeV. In all the nuclei considered the excitation energy of the lowest $2^+$ state is rather well reproduced. Moreover, the $B(E2)$ transition probability from the ground state to the $2^+$ state is also well reproduced in the three cases. Other quantities like two neutron and two proton separation energies also compare well with experimental data.

**Acknowledgments**

One of us (R. R.-G.) kindly acknowledges the financial support received from the Spanish Instituto de Cooperación Iberoamericana (ICI). This work has been supported in part by the DGICyT (Spain) under project PB97/0023.